 \documentclass[preprint2]{aastex}    
\usepackage{natbib}
\usepackage{bm}
\usepackage{array}
 \usepackage{verbatim}

\usepackage{gensymb}
\usepackage{graphicx,amssymb,amstext,amsmath}
\usepackage{threeparttable}
\usepackage{color}

\bibpunct[, ]{(}{)}{;}{a}{}{,}
\newcommand{\hMpc}{{\ifmmode{h^{-1}{\rm Mpc}}\else{$h^{-1}$Mpc }\fi}}
\newcommand{\hkpc}{{\ifmmode{h^{-1}{\rm kpc}}\else{$h^{-1}$kpc }\fi}}
\newcommand{\hMsun}{{\ifmmode{h^{-1}{\rm {M_{\odot}}}}\else{$h^{-1}{\rm{M_{\odot}}}$}\fi}}
\newcommand{\Msun}{{\ifmmode{{\rm {M_{\odot}}}}\else{${\rm{M_{\odot}}}$}\fi}}

\begin{document}

\title{
COSMOLOGICAL PARAMETERS FROM SUPERNOVAE ASSOCIATED WITH GAMMA-RAY BURSTS}
\author{Xue Li\altaffilmark{1}, Jens Hjorth\altaffilmark{1} and Rados{\l}aw Wojtak\altaffilmark{1,2}}
\altaffiltext{1}{Dark Cosmology Centre, Niels Bohr Institute, University of Copenhagen, Juliane Maries Vej 30, DK-2100 Copenhagen, Denmark}
\email{lixue@dark-cosmology.dk}
\altaffiltext{2}{Kavli Institute for Particle Astrophysics and Cosmology, Stanford University,
SLAC National Accelerator Laboratory, Menlo Park, CA 94025}

\begin{abstract}
We report estimates of the cosmological parameters $\Omega_m$ and 
$\Omega_{\Lambda}$ obtained using supernovae (SNe) associated with
gamma-ray bursts (GRBs) at redshifts up to 0.606.    Eight 
high-fidelity GRB-SNe with well-sampled light curves across the peak are used. 
We correct their peak 
magnitudes for a luminosity-decline rate relation to turn them into accurate 
standard candles with dispersion  $\sigma = 0.18$  mag.   We also estimate the peculiar velocity of the low-redshift  host galaxy of SN 1998bw, using constrained cosmological simulations.          In a flat universe, the resulting Hubble diagram leads to 
  best-fit cosmological parameters of    $(\Omega_m,  \Omega_{\Lambda}) =  
 (0.58^{+0.22}_{-0.25},0.42 ^{+0.25}_{-0.22})$.    
This exploratory study suggests that GRB-SNe can potentially be used as 
standardizable candles to high redshifts to measure distances in the universe and 
constrain cosmological parameters.  
 
\end{abstract}
 


\section{INTRODUCTION}

The accelerating expansion of the universe was detected with the help of  
Type Ia supernovae (SNe Ia)  \citep{perlmutter_1997, riess,  perlmutter}. Taking advantage
of the correlation between  their decline rate  and peak brightness 
\citep{phillips_1993, phillips_1999}, the corrected    luminosities
of SNe Ia exhibit sufficiently small dispersons that they can be
used to measure cosmological distances and constrain cosmological parameters.

While SNe Ia  are exquisite standard candles that are routinely used
to measure distances out to $z \approx 1.0$ \citep{ koekemoer_2011,hook_2013}, the rate of events unfortunately appears to decline
at higher redshifts \citep{graur_2013, rodney_snr}. However, it is
necessary to observe the universe  at redshift $z>1$ to constrain the
dark-energy equation of state parameter $w(z)$  by  breaking the 
degeneracies between cosmological models 
\citep{linder_2003, king_2013}.  

At  higher redshifts, e.g.,  $z>1.5$, core collapse supernovae (CCSNe) strongly 
dominate the rates of SNe \citep{lenspaper,rodney_snr}.     Moreover, with  more powerful telescopes to be launched, e.g., the \emph{James Webb 
Space Telescope} (\emph {JWST}), CCSNe may be discovered at   
redshifts up to $z = 7-8$ \citep{pan_2013}. But in general, CCSNe are much 
fainter than SNe Ia.   They do not have the same intrinsic luminosities and their peak magnitudes do not exhibit any correlation with the decline rates  \citep{drout_ibc}. 

This problem may be solved by considering a certain type of CCSNe: a subclass of broad-lined Type Ic SNe which are observed to be associated with 
gamma-ray bursts (GRBs).  First observed by the Vela Satellites in 1967   \citep{klebesadel_1973},   GRBs are   flashes of narrow beams of intense electromagnetic radiation whose peak energies occur at gamma-ray wavelengths \citep{metzger_1997}.   SN 1998bw was first detected to be associated with GRB 980425  \citep{galama_1998bw, iwamoto_1998, kulkarni_1998, woosley_1998}. Since then,  many GRB-SNe have been found \citep{hjorth_2003dh, stanek_2003dh, woosley_2006, hjorth_connection}.

GRB-SNe have relatively smooth optical spectra and very large 
explosion energies \citep{galama_1998bw, hjorth_2013}.  
The peak magnitudes of GRB-SNe
are in the same range as SNe Ia.   Moreover,   their  peak magnitudes are correlated with their decline rates  \citep{lixue_2013sn}. While GRB-SNe
are rare and difficult to disentangle from the contaminating light of
the GRB afterglow and host galaxy, these properties could make GRB-SNe a 
powerful tool for distance determination and constraining cosmological parameters.  This paper is devoted to the first quantitative exploration of this idea.

The outline of the  paper is as follows. In Section  \ref{sec:data},  
we briefly review the procedure of obtaining  
 light curves of  GRB-SNe, and 
 measuring peak magnitudes and decline rates.  We also present   an estimate of the peculiar velocity of  the host galaxy of the low-redshift SN 1998bw. 
In Section \ref{sec:candle},  we 
establish GRB-SNe as standard candles
based on a limited set of high-quality GRB-SNe.
In  Section \ref{sec:cos} we create a Hubble diagram  
and place constraints on the matter density parameter assuming a flat $\Lambda$CDM cosmological model. 
We     conclude in  Section \ref{sec:summary}.

 
 \section {GRB-SNE SYSTEMS}
 \label{sec:data}

\begin{table*}[h!tbp]
\begin{center}
\caption{ Light curve properties of GRB-SNe systems      }  
 \label{listgrbsn}
 \begin{tabular}{ccccc} 
  
\hline       
\hline   
  GRB/XRF/SN &    $z$ &  $m_{V}^{\rm corr} $ $^a$      &  $\Delta m_{V, 15}$ $^b$           \\ 
                             &             &      (mag)                                &  (mag)                 &                       \\
   \hline

980425/1998bw  &   0.00857    $^c$  & 13.66$^{+0.08}_{-0.08}$ & $0.75^{+0.02}_{-0.02}$        \\
030329/2003dh  & 0.1685  & 20.23$^{+0.15}_{-0.12}$ & $0.90^{+0.50}_{-0.50}$       \\
031203/2003lw  & 0.1055  & 18.62$^{+0.16}_{-0.16}$ & $0.64^{+0.10}_{-0.10}$      \\
050525A/2005nc & 0.606   & 24.24$^{+0.39}_{-0.23}$ & $1.17^{+0.77}_{-0.85}$         \\
060218/2006aj  & 0.03342 & 17.07$^{+0.08}_{-0.08}$ & $1.09^{+0.06}_{-0.06}$      \\
090618         & 0.54    & 23.20$^{+0.13}_{-0.13}$ & $0.67^{+0.16}_{-0.19}$      \\
100316D/2010bh & 0.059   & 18.31$^{+0.10}_{-0.10}$ & $1.10^{+0.05}_{-0.05}$      \\
120422A/2012bz & 0.283   & 21.40$^{+0.03}_{-0.03}$ & $0.73^{+0.06}_{-0.06}$    \\

  \hline
                                                                   
\end{tabular}
\end{center}
{\footnotesize
 \noindent
 $^a$: Here   $m_{V}^{\rm corr} $  is the apparent magnitude after extinction correction   and K correction    \citep{lixue_2013sn}.      \\
$^b$: Here $\Delta m_{V, 15}$ represents    the decline of the  rest frame V-band magnitude 15 days   after the SN   reaches its peak brightness.  \\
$^c$:     More details are in Section \ref{sub:pec}. \\
 }
\end{table*}

 The selected GRB-SNe   are   firmly associated with GRBs         from class  {\it A} to class {\it C}      \citep{hjorth_connection},  where class {\it A}  has   `strongest spectroscopic evidence'.    Here we briefly summarize the discussion of  the steps in obtaining the light curves of GRB-SNe.  More details on the procedure   are in  \cite{lixue_2013sn}.   The systems  are listed in Table \ref{listgrbsn}.

 \subsection{Data  Analysis}

The afterglow is either fitted to power-law or broken power-law functions and subtracted.    Both Galactic and host extinction are corrected for.  For the Galactic extinction,   
we assume  $R_V = 3.1$ and  get $E(B-V)$ from  the  DIRBE/IRAS   dust map \citep{schlegel_dustmap}.   The values are re-calibrated   based on  Table 6  in  \cite{ schlafly_dustmap}.      
We take the values of host extinction  from the literature.   
We fit low-order polynomial functions to obtain  the light curves.   A K correction method is developed to correct the peak magnitudes and decline rates into the rest frame V band. A `multi-band K-correction' is used for   systems which have two band data available and these two bands are close to the redshifted V band.  Otherwise, SN 1998bw peak SED and decline rate templates are used to correct the peak magnitude  and the decline rate  from the light curve  obtained at a wavelength  close  to the redshifted   V band.    In total eight  light curves of  GRB-SNe with $z$ up to 0.606 are obtained   in the rest frame V band \citep{lixue_2013sn}.

 \subsection{Peculiar Velocity and Uncertainty of Distance Modulus of SN 1998bw}
 \label{sub:pec}

 
\begin{figure}
\centerline{\includegraphics[width = 8.8 cm]{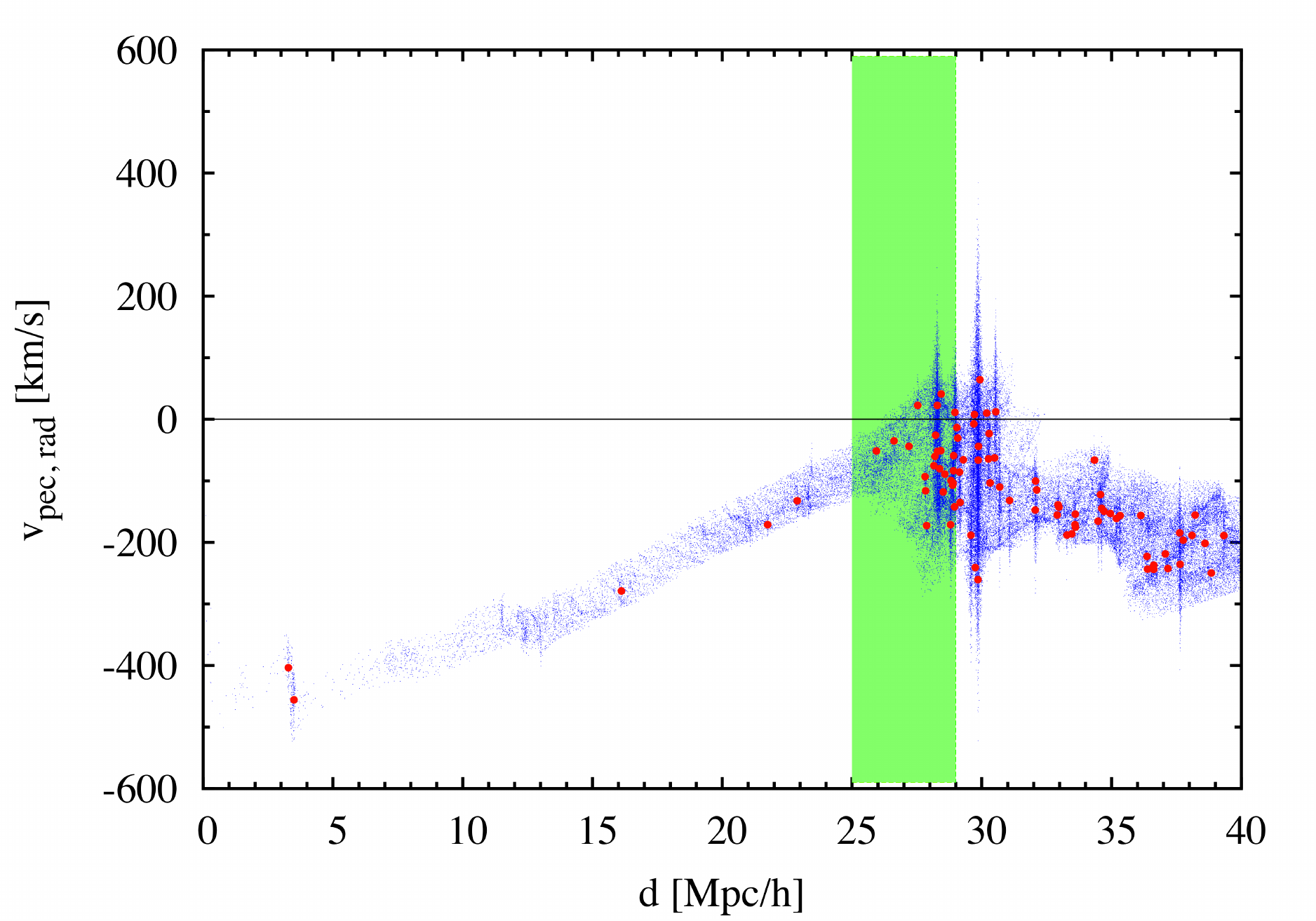}}
\caption{Peculiar velocities (line-of-sight component) in the direction of the   supernova host galaxy,   ESO 184$-$G82, as 
a function of the comoving distance from the Local Group formed the constrained simulation. 
The blue dots show dark matter particles and the red symbols represent dark matter haloes. The 
green band indicates the location of  the galaxy ESO 184$-$G82. }
  \label{pec}
\end{figure}

SN 1998bw \citep{galama_1998bw, iwamoto_1998, kulkarni_1998, woosley_1998} was the first SN discovered to be connected with a GRB (GRB 980425).    It is the nearest    GRB-SN  so far, and the measured redshift 
is    $z = 0.00867 \pm 0.00004$  \citep{foley_2006}, so it      constitutes  an important low-redshift anchor of the Hubble diagram.        For the recessional velocity of this low-redshift system,  the contribution from the peculiar velocity may be relatively substantial.   
The true recessional velocity $v_{\rm rec}$ due to the Hubble flow should be corrected for the peculiar velocity $v_{\rm pec}$ of the host galaxy: $v_{\rm CMB} = v_{\rm rec}  + v_{\rm pec}$, with  $v_{\rm CMB}$ being the velocity relative to the cosmic microwave background (CMB).

In order to estimate the peculiar velocity of  its  host galaxy,   ESO 184$-$G82,   and    calculate the uncertainty in the distance modulus,       we   use a dark matter simulation performed 
as a part of the Constrained Local Universe Simulations (CLUES) project.  
The simulation is carried out in a 
volume of $(160 h^{-1} {\rm Mpc})^3$ containing $1024^{3}$ particles. The assumed cosmological model is based on 
the 3rd data release of  the WMAP satellite (WMAP3 cosmology), i.e.,   matter 
density $\Omega_{m}=0.24$,   dimensionless Hubble parameter $h=0.73$,  and   normalization of the power 
spectrum $\sigma_{8}=0.76$. The initial conditions are generated from observational data of the 
galaxy distribution and galaxy velocities in the local universe \citep[for technical details see][]{Got2010}. With this setup, the simulation 
recovers all observed structures on scales larger than $5 h^{-1} {\rm Mpc}$. In particular, all nearby galaxy clusters and superclusters, 
such as the Virgo cluster, the Coma cluster, the Great Attractor and the Perseus--Pisces cluster, are well reproduced 
in the final simulation snapshot. On the other hand, small scale structures formed in the simulation emerge from a random 
realization of the power spectrum on these scales. Their evolution, however, is strongly constrained by nearby large-scale 
structures. Therefore, the simulation provides a realistic and dynamically self-consistent model for the matter distribution 
and the velocity field in the local universe.

The position vector of the host galaxy with respect to the Local Group in the simulation box can be found be matching 
angular separations from several large-scale structures. As the reference structures, we use  the Coma cluster, the 
Perseus--Pisces cluster and the Great Attractor.  Having determined the direction to the host galaxy from the Local Group 
in the simulation box, we compute  the radial components of peculiar velocities within a narrow light cone. Figure~\ref{pec} shows 
the resulting projected peculiar velocities as a function of the  comoving distances from the Local Group. The blue dots 
show velocities of dark matter particles, whereas the red symbols represent dark matter haloes found with the 
friends-of-friends  algorithm. Lack of dark matter haloes at small distances is related to the fact that the line 
of sight crosses the edge of the Local Void \citep[see][]{Nas2011}.

  To identify the position of the host galaxy  ESO 184$-$G82 in the simulation box, we use a range of plausible distances to the host galaxy in units of Mpc/h so they are independent of $H_0$.  We assume that the true recessional velocity is likely  between the host velocity with respect to the CMB   $v_{\rm CMB}  = 2505 \pm 14$  km s$^{-1}$ \citep{foley_2006} and the host velocity   with respect to the  local large-scale structures (Virgo,  Great Attractor  and Shapley Supercluster) $v_{\rm Virgo + GA + Shapley} = 2769 \pm 21$  km s$^{-1}$, as shown in the green band in  Figure~\ref{pec} \footnote{more details are in http://ned.ipac.caltech.edu/ and the reference therein}.  Within the green band,  the mean peculiar velocity is   $v_{\rm pec}  = -65$  km s$^{-1}$  with a mean systematic error   $\pm 75$  km s$^{-1}$.   Combined with the peculiar velocity   $v_{\rm pec}$   and the CMB velocity $ v_{\rm CMB} $,  the Hubble flow velocity is   $v_{\rm rec}  = 2570 \pm 76$  km s$^{-1}$, which is in the green band in  Figure~\ref{pec},  as expected.   The uncertainty of $ v_{\rm pec}$ dominates the uncertainty of $v_{\rm rec}$. 
The corresponding  redshift is  
  $z = v_{\rm rec}  / c = 0.00857 \pm 0.00025$.  Therefore, the contribution of  the peculiar velocity  of the  the host galaxy to uncertainty in the distance modulus of SN 1998bw is $\sigma ({\rm DM}) = (5/2.3)  \sigma (v_{\rm pec})  (cz)^{-1} =  0.06$ mag, where $c$ is the speed of light and $ \sigma (v_{\rm pec})$ is the uncertainty of peculiar velocity $v_{\rm pec}$.

\section{GRB-SNE AS  STANDARD CANDLES}
\label{sec:candle}

    In much the same way as   SNe Ia  were used   to measure  the  
 cosmological parameters  $\Omega_m$  
 and $\Omega_{\Lambda}$  \citep{perlmutter_1997, riess,  perlmutter}, here  
 we use GRB-SNe as standard candles.  

 Similar to SNe Ia  \citep{phillips_1993, phillips_1999}, 
  GRB-SNe   have     bright  peak luminosities.   The luminosity-decline rate relation  for GRB-SNe  in the rest-frame V band  is     \citep{lixue_2013sn} 
        \begin{equation} 
 M_{\rm V,peak} = \alpha  \Delta m_{V, 15} +  M_0,    
  \label{eq;relation}
\end{equation} 
 where $\alpha$  is the slope  and   $M_0$ is  a constant representing the absolute peak magnitude      at  $ \Delta m_{V, 15} = 0$.     
  Assuming   $\Omega_m = 0.315$ and $H_0 = 67.3$  km s$^{-1}$ Mpc$^{-1}$  \citep{planck_cosmopara},    we have  $\alpha = 1.57^{+0.25}_{-0.28}$  and  $M_0  = - 20.58^{+0.22}_{-0.20}$  \citep{lixue_2013sn}.    
 This relation 
  is superior to other similar relations \citep{lixue_2013sn}, such as the  $k-s$ relation  \citep{cano_ks2014},  where $k$ and $s$ are the relative peak and width of the light curves compared to SN 1998bw, or the relation between the peak magnitude and the elapsed  time since GRB.     With  $ M_{\rm V,peak}$  from  \cite{lixue_2013sn},  obtained using the above Planck cosmology,    the corrected apparent peak magnitude  in the rest-frame V band is  
   \begin{equation} 
m_{V}^{\rm corr}  =          M_{\rm V,peak} + DM(z),     
 \label{eq:obsmag}  
\end{equation}  
where $m_{V}^{\rm corr} $ is the corrected apparent magnitude in the rest-frame V band after   corrections    for    dust extinction   and K correction  \citep{lixue_2013sn}.     
Here   $DM(z)$     
is the distance modulus at $\Omega_m = 0.315$ and $H_0 = 67.3$  km s$^{-1}$ Mpc$^{-1}$ in a flat universe.   The values of   $m_{V}^{\rm corr} $ are listed in  Table  \ref{listgrbsn}.     
Considering the relation  in  Eq. (\ref{eq;relation}),  
 the effective apparent  peak magnitude    $m_V^{\rm eff} $ can be obtained as
    \begin{equation} 
  m_V^{\rm eff}  =  m_V^{\rm corr} -  \alpha \Delta m_{V, 15}.   
  \label{effectivemag}
\end{equation}  
The term  $\alpha \Delta m_{V, 15}$   represents the correction due to the  luminosity-decline rate    relation.  The effective apparent magnitude can  also be expressed as  \citep{perlmutter_1997, perlmutter} 
 \begin{equation} 
 m_V^{\rm eff}  =   \Upsilon + 5 \log  \mathcal{D}_{L}(z; \Omega_m, \Omega_{\Lambda}), 
  \label{magandcosmology}
 \end{equation} 
where $\mathcal{D}_{L} \equiv H_0 d_L$ is the \textquotedblleft $H_0$-free\textquotedblright\ luminosity distance in units of km s$^ {-1}$, with   $d_L$ being the luminosity distance in units of Mpc \citep{hogg_distance} and  $H_0$ in units of km s$^{-1}$ Mpc$^{-1}$. Here  $\Upsilon = M_0 -  5 \log H_0 + 25$    is the \textquotedblleft $H_0$-free\textquotedblright\ V-band absolute peak magnitude         \citep{perlmutter_1997, perlmutter}.    
The fitting procedure does not invoke $H_0$  and the  constraints on cosmological parameters are therefore independent of the Hubble constant.    
In this paper,   $\alpha$ and $\Upsilon$ are statistical `nuisance' parameters.

 \section{CONSTRAINTS ON $\Omega_m$ and $\Omega_{\lambda}$}
 
 \label{sec:cos}
 
 
\begin{figure}[htbp]
\centerline{\includegraphics[width = 8.8 cm]{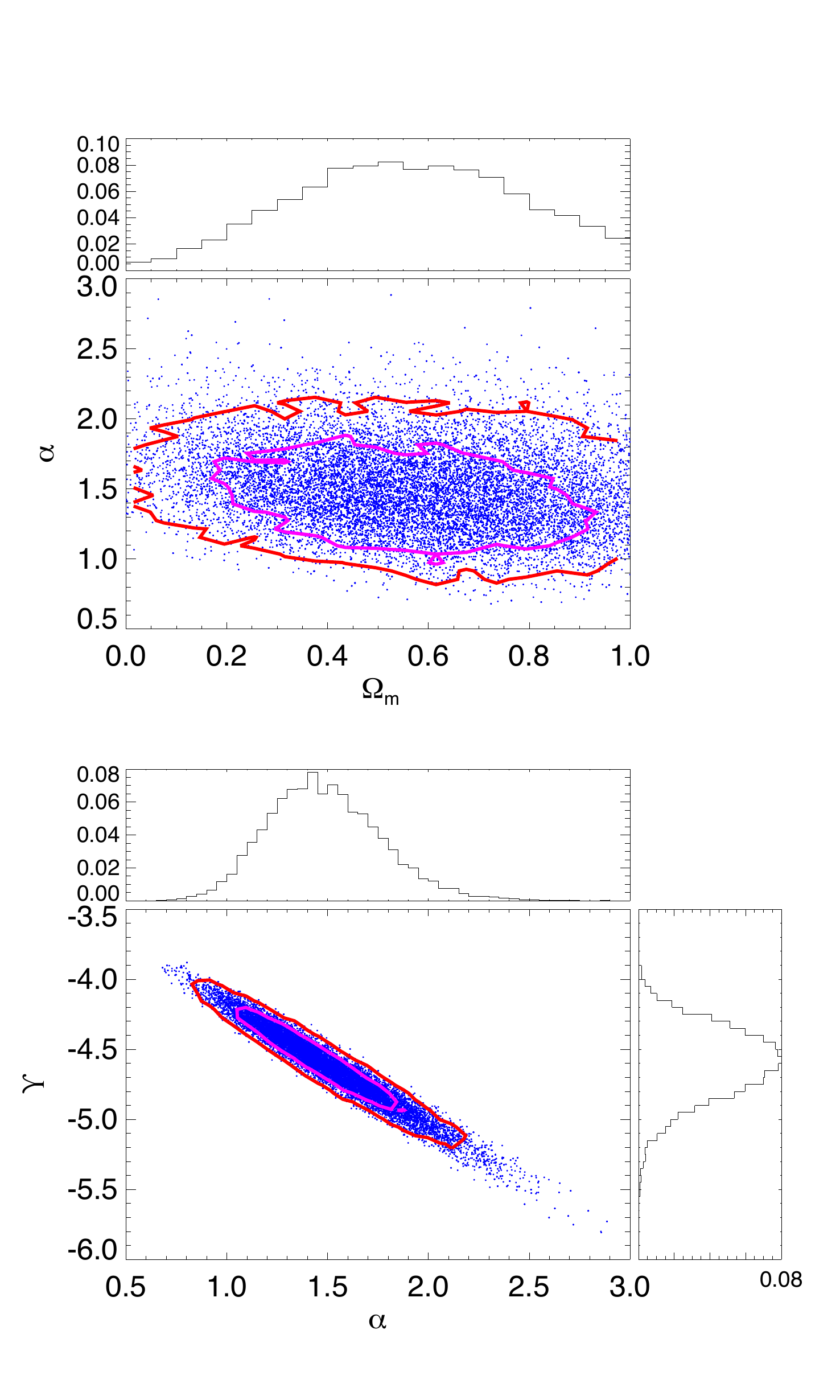}}
\caption{	Constraints on  
$\alpha$,  $\Upsilon$   
and $\Omega_m$   assuming a flat cosmological model.    The confidence levels  of the density contours  are 68.3\%  and 95.5\%.     
  }
  \label{fig:para}
\end{figure}

 
\begin{figure}[htbp]
\centerline{\includegraphics[width = 8.8 cm]{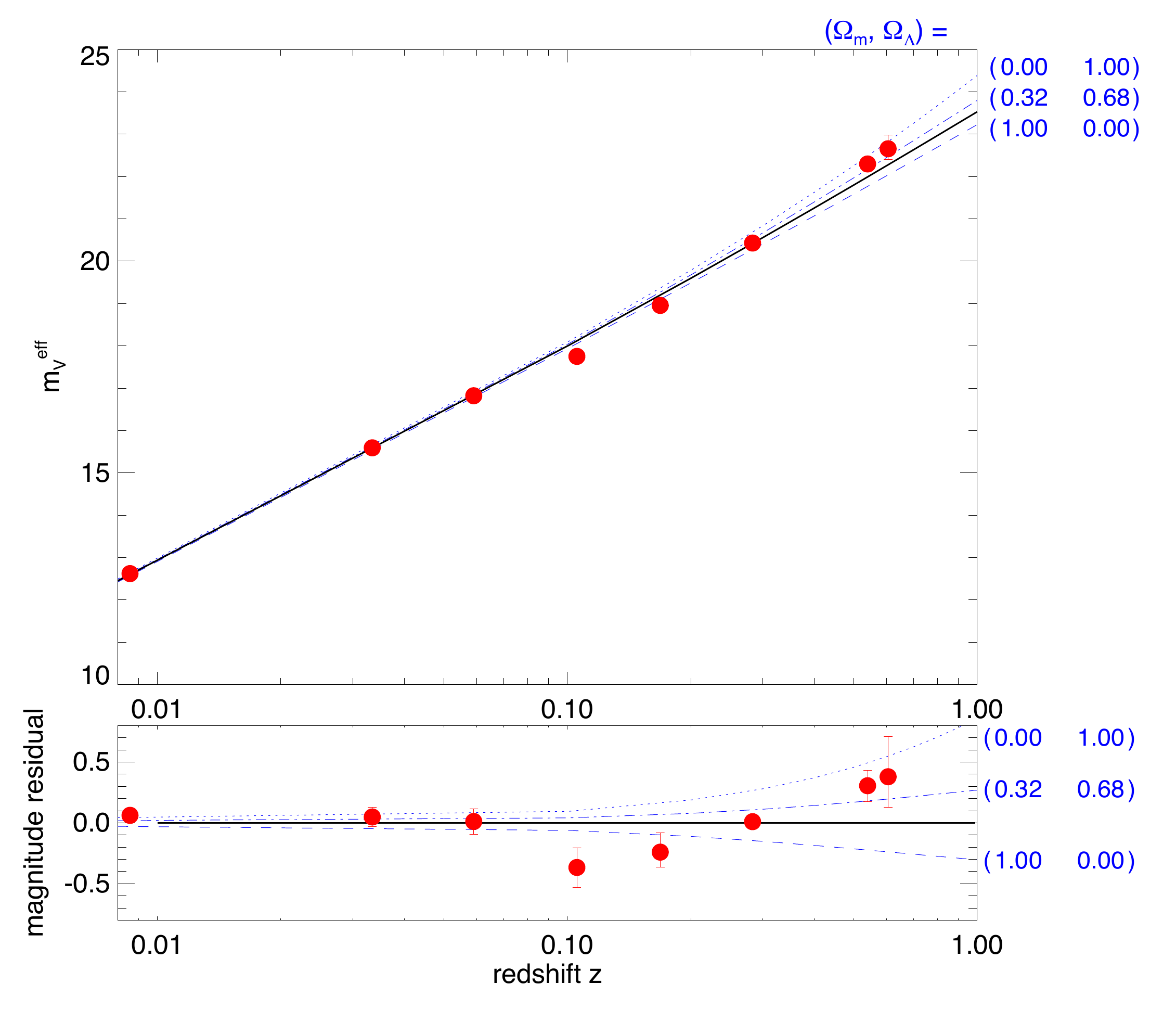}}
\caption{	 Hubble diagram for   GRB-SNe.  
The effective apparent magnitudes of eight GRB-SNe plotted as red points,   are calculated using 
   the best-fit  parameters $\alpha = 1.37$ and $\Upsilon =  - 4.50$.     The best cosmological model with $(\Omega_m,  \Omega_{\Lambda}) =  
(0.58 ,0.42)$  is   in black, while other models (labeled on the right side of the figure) are in blue.     The lower panel shows the   magnitude residuals from the best-fit cosmological model.    }
  \label{fig:hd}
\end{figure}


  We employ a Monte Carlo Markov Chain technique to place constraints on the matter density parameters 
and the two nuisance parameters. We adopt a flat cosmological model, 
i.e. $\Omega_{m}+\Omega_{\Lambda}=1$, and assume a flat prior on all free parameters, i.e. $\Omega_{m}$, 
$\alpha$ and $\Upsilon$. 
With a straightforward generalization of the $\chi^2$ 
function from  \cite{snls_2005}, the adopted likelihood function $L$ is 
 \begin{eqnarray}
L\propto \prod_{i} \exp\Big[ \frac{\Delta_{i}^2}{2\sigma_{i}^{2}} \Big]\frac{1}{\sigma_{i}},
\end{eqnarray}
with $\Delta_i = m_{V,i}^{\rm corr}-\alpha\Delta m_{V,15,i}-\Upsilon-5\log \mathcal{D}_{L}(z_{i},\Omega_{m})$, where $\sigma_{i}^{2}=\sigma^{2}(m_{V,i}^{\rm corr})+\alpha^{2}\sigma^{2}(\Delta m_{V,15,i})$, and $\sigma(m_{V,i}^{\rm corr})$ and $\sigma(\Delta m_{V, 15,  i})$ are  
the errors of $m_{V,i}^{\rm corr}$ and $\Delta m_{V, 15, i}$,   respectively. The formula for $\sigma_{i}$ assumes an independent propagation of 
errors in $m_{V,i}^{\rm corr}$ and $\Delta m_{V,15, i}$. We verify this assumption by finding no signature of a correlation between $m_{V}^{\rm corr}$ and 
$\Delta m_{V,15}$ in a covariance matrix obtained from fitting the light curves.

  The   marginalised posterior probability densities of $\Omega_m$, $\alpha$ and $\Upsilon$  are shown in Figure \ref{fig:para}.    
 The  confidence levels of the density contours  are 68.3\%  and 95.5\%.  
We quantify the fit in terms of the maximum-likelihood values and confidence intervals containing 68.3\% of the corresponding marginal probabilities.     
The best-fit nuisance parameters are  $\alpha   = 1.37 ^{+0.36 }_{-0.19 }$ and $\Upsilon  =  -4.50^{+0.17}_{-0.32}$,   which  are        consistent with the  values derived from  the luminosity-decline rate relation,     
assuming    
   $\Omega_m = 0.315$ and $H_0 = 67.3$  km s$^{-1}$ Mpc$^{-1}$  \citep{lixue_2013sn}.         In  a flat universe, the best-fit  cosmological model is   $(\Omega_m,  \Omega_{\Lambda}) =  
(0.58^{+0.22}_{-0.25},0.42 ^{+0.25}_{-0.22})$.      


Figure  \ref{fig:hd} shows the Hubble diagram  for eight  GRB-SNe systems. The effective magnitudes  $m_V^{\rm eff}$ of the GRB-SN systems in the rest frame V  band     are plotted as red points.   The best cosmological model  is the black curve. For comparison, we plot   three other  cosmological models: \{$(\Omega_m,  \Omega_{\Lambda})$\}  $= \{(0, 1),   (0.32, 0.68),   (1, 0)\}$ as dotted lines.  The lower panel in Figure  \ref{fig:hd} shows the magnitude residuals relative to the best cosmological model.

\section {CONCLUSION}
\label{sec:summary}

   The cosmological parameters $\Omega_m$ and $\Omega_{\Lambda}$ can be constrained with SNe Ia  \citep{perlmutter,knop_2003}, CMB radiation \citep{spergel_2003,planck_cosmopara}, and clusters of galaxies \citep{allen_2002,jullo_2010}.    
   As shown in this paper, GRB-SNe may add further constraints
on cosmological parameters.     With more systems at  $z$ up to 1,  the result would be more constraining, but we have opted here for systems with very well sampled light curves.        
At higher redshifts ($z>1.5$),  with more powerful telescope  to be launched,  e.g.,  \emph {JWST},   GRB-SNe are potential candidates to break the degeneracies and constrain the equation of state parameter  $w(z)$ \citep{linder_2003, king_2013}.


\acknowledgments

 We thank  Enrico Ramirez-Ruiz, Tomotsugu Goto   and Dong Xu   for their many helpful discussions.   We thank Stefan Gottl\"ober for making available one of the CLUES simulations ({\tt http://www.clues-project.org/}). The simulation has been performed at the 
Leibniz Rechenzentrum (LRZ), Munich.   The Dark Cosmology Centre is funded by the Danish National Research Foundation.    

    

\clearpage
\newpage


\begin{thebibliography}{49}
\expandafter\ifx\csname natexlab\endcsname\relax\def\natexlab#1{#1}\fi


\bibitem[{{Allen}  {et~al.}(2002),  {Allen} {Schmidt} \& {Fabian}}]{allen_2002}
               {Allen}, S.~W.,   {Schmidt}, R.~W.    \& {Fabian}, A.~C.  2002,  \mnras, 334, L11
  
\bibitem[{{Astier}  {et~al.}(2006) {Astier}, P. and {Guy}, J. and {Regnault}, N. and {Pain}, R. and 
	{Aubourg}, E. and {Balam}, D. and {Basa}, S. and {Carlberg}, R.~G. and 
	{Fabbro}, S. and {Fouchez}, D. and {Hook}, I.~M. and {Howell}, D.~A. and 
	{Lafoux}, H. and {Neill}, J.~D. and {Palanque-Delabrouille}, N. and 
	{Perrett}, K. and {Pritchet}, C.~J. and {Rich}, J. and {Sullivan}, M. and 
	{Taillet}, R. and {Aldering}, G. and {Antilogus}, P. and {Arsenijevic}, V. and 
	{Balland}, C. and {Baumont}, S. and {Bronder}, J. and {Courtois}, H. and 
	{Ellis}, R.~S. and {Filiol}, M. and {Gon{\c c}alves}, A.~C. and 
	{Goobar}, A. and {Guide}, D. and {Hardin}, D. and {Lusset}, V. and 
	{Lidman}, C. and {McMahon}, R. and {Mouchet}, M. and {Mourao}, A. and 
	{Perlmutter}, S. and {Ripoche}, P. and {Tao}, C. and {Walton}, N.
	}]   {snls_2005}
	{Astier}, P.,   {et~al.}   2006, \aap,  447, 31

\bibitem[{{Cano}(2014)}]{cano_ks2014}
                          {Cano}, Z. 2014, accepted by \apj  
                          
\bibitem[{{Drout}   {et~al.}(2011) {Drout}, M.~R. and {Soderberg}, A.~M. and {Gal-Yam}, A. and 
	{Cenko}, S.~B. and {Fox}, D.~B. and {Leonard}, D.~C. and {Sand}, D.~J. and 
	{Moon}, D.-S. and {Arcavi}, I. and {Green}, Y.}]{drout_ibc}
                  {Drout}, M.~R.,    {et~al.}   2011, \apj, 741,  97 


\bibitem[{{Foley} {et~al.}(2006) {Foley}, S. and {Watson}, D. and {Gorosabel}, J. and {Fynbo}, J.~P.~U. and 
	{Sollerman}, J. and {McGlynn}, S. and {McBreen}, B. and {Hjorth}, J.}] {foley_2006}
         {Foley}, S.,  {Watson}, D.,  {Gorosabel}, J.,  {Fynbo}, J.~P.~U.,  
	{Sollerman}, J.,  {McGlynn}, S.,  {McBreen}, B. \&   {Hjorth}, J.  2006, \aap,  447, 891
	
	
\bibitem[{{Galama} {et~al.}(1998) {Galama}, T.~J. and {Vreeswijk}, P.~M. and {van Paradijs}, J. and 
	{Kouveliotou}, C. and {Augusteijn}, T. and {B{\"o}hnhardt}, H. and 
	{Brewer}, J.~P. and {Doublier}, V. and {Gonzalez}, J.-F. and 
	{Leibundgut}, B. and {Lidman}, C. and {Hainaut}, O.~R. and {Patat}, F. and 
	{Heise}, J. and {in't Zand}, J. and {Hurley}, K. and {Groot}, P.~J. and 
	{Strom}, R.~G. and {Mazzali}, P.~A. and {Iwamoto}, K. and {Nomoto}, K. and 
	{Umeda}, H. and {Nakamura}, T. and {Young}, T.~R. and {Suzuki}, T. and 
	{Shigeyama}, T. and {Koshut}, T. and {Kippen}, M. and {Robinson}, C. and 
	{de Wildt}, P. and {Wijers}, R.~A.~M.~J. and {Tanvir}, N. and 
	{Greiner}, J. and {Pian}, E. and {Palazzi}, E. and {Frontera}, F. and 
	{Masetti}, N. and {Nicastro}, L. and {Feroci}, M. and {Costa}, E. and 
	{Piro}, L. and {Peterson}, B.~A. and {Tinney}, C. and {Boyle}, B. and 
	{Cannon}, R. and {Stathakis}, R. and {Sadler}, E. and {Begam}, M.~C. and 
	{Ianna}, P.}] {galama_1998bw}
      {Galama}, T.~J.,  {et~al.} 1998, \nat, 395, 670	


\bibitem[{{Gottloeber} {et~al.}(2010) {Gottloeber}, S. and {Hoffman}, Y. and {Yepes}, G.}] {Got2010}
                 {Gottloeber}, S.,    {Hoffman}, Y.,  \& {Yepes}, G. 2010, ArXiv e-prints 
 
 
\bibitem[{{Graur} {et~al.}(2014) {Graur}, O.   and {Rodney}, S.~A. and {Maoz}, D. and {Riess}, A.~G. and 
	{Jha}, S.~W. and {Postman}, M. and {Dahlen}, T. and {Holoien}, T.~W.-S. and 
	{McCully}, C. and {Patel}, B. and {Strolger}, L.-G. and {Ben{\'{\i}}tez}, N. and 
	{Coe}, D. and {Jouvel}, S. and {Medezinski}, E. and {Molino}, A. and 
	{Nonino}, M. and {Bradley}, L. and {Koekemoer}, A. and {Balestra}, I. and 
	{Cenko}, S.~B. and {Clubb}, K.~I. and {Dickinson}, M.~E. and 
	{Filippenko}, A.~V. and {Frederiksen}, T.~F. and {Garnavich}, P. and 
	{Hjorth}, J. and {Jones}, D.~O. and {Leibundgut}, B. and {Matheson}, T. and 
	{Mobasher}, B. and {Rosati}, P. and {Silverman}, J.~M. and {U}, V. and 
	{Jedruszczuk}, K. and {Li}, C. and {Lin}, K. and {Mirmelstein}, M. and 
	{Neustadt}, J. and {Ovadia}, A. and {Rogers}, E.~H.}] {graur_2013}
          {Graur}, O.,  {et~al.}   2014, \apj, 783, 28
	
\bibitem[{{Hjorth}(2013) {Hjorth}, J.}]{hjorth_2013}
                {Hjorth}, J. 2013,  Royal Society of London Philosophical Transactions Series A, 371, 20275
                
\bibitem[{{Hjorth} \& {Bloom}(2012)  {Hjorth}, J. and {Bloom}, J.~S.}] {hjorth_connection}
                   {Hjorth}, J., \& {Bloom}, J.~S.  2012,  The Gamma-Ray Burst - Supernova Connection, 169-190
                   


\bibitem[{{Hjorth} {et~al.}(2003) {Hjorth}, J. and {Sollerman}, J. and {M{\o}ller}, P. and {Fynbo}, J.~P.~U. and 
	{Woosley}, S.~E. and {Kouveliotou}, C. and {Tanvir}, N.~R. and 
	{Greiner}, J. and {Andersen}, M.~I. and {Castro-Tirado}, A.~J. and 
	{Castro Cer{\'o}n}, J.~M. and {Fruchter}, A.~S. and {Gorosabel}, J. and 
	{Jakobsson}, P. and {Kaper}, L. and {Klose}, S. and {Masetti}, N. and 
	{Pedersen}, H. and {Pedersen}, K. and {Pian}, E. and {Palazzi}, E. and 
	{Rhoads}, J.~E. and {Rol}, E. and {van den Heuvel}, E.~P.~J. and 
	{Vreeswijk}, P.~M. and {Watson}, D. and {Wijers}, R.~A.~M.~J.}] {hjorth_2003dh}
	 {Hjorth}, J.,  {et~al.}  2003, \nat, 423, 847 
	
\bibitem[{{Hogg}(1999) {Hogg}, D.~W.}]{hogg_distance}
                 {Hogg}, D.~W. 1999, ArXiv Astrophysics e-prints

\bibitem[{{Hook}(2013)  {Hook}, I.~M.}]  {hook_2013}
                    {Hook}, I.~M.  2013,  Royal Society of London Philosophical Transactions Series A, 371, 20282

\bibitem[{{Iwamoto}  {et~al.}(1998) {Iwamoto}, K. and {Mazzali}, P.~A. and {Nomoto}, K. and {Umeda}, H. and 
	{Nakamura}, T. and {Patat}, F. and {Danziger}, I.~J. and {Young}, T.~R. and 
	{Suzuki}, T. and {Shigeyama}, T. and {Augusteijn}, T. and {Doublier}, V. and 
	{Gonzalez}, J.-F. and {Boehnhardt}, H. and {Brewer}, J. and 
	{Hainaut}, O.~R. and {Lidman}, C. and {Leibundgut}, B. and {Cappellaro}, E. and 
	{Turatto}, M. and {Galama}, T.~J. and {Vreeswijk}, P.~M. and 
	{Kouveliotou}, C. and {van Paradijs}, J. and {Pian}, E. and 
	{Palazzi}, E. and {Frontera}, F.}] {iwamoto_1998} 
	{Iwamoto}, K.,   {et~al.}  1998, \nat, 395, 672 
	
\bibitem[{{Jullo} {et~al.}(2010)  {Jullo}, E. and {Natarajan}, P. and {Kneib}, J.-P. and {D'Aloisio}, A. and 
	{Limousin}, M. and {Richard}, J. and {Schimd}, C.}]     {jullo_2010}
	{Jullo}, E., {Natarajan}, P., {Kneib}, J.-P., {D'Aloisio}, A., 
	{Limousin}, M.,  {Richard}, J.,  \&  {Schimd}, C.   2010, Science, 329, 924 

\bibitem[{{King}  {et~al.}(2013) {King}, A.~L. and {Davis}, T.~M. and {Denney}, K. and {Vestergaard}, M. and 
	{Watson}, D.}] {king_2013}
	 {King}, A.~L.,  {Davis}, T.~M., {Denney}, K., {Vestergaard}, M., \&  
	{Watson}, D.   2013, \mnras,  441, 3454-3476
	
	
\bibitem[{{Klebesadel}  {et~al.}(1973) {Klebesadel}, R.~W. and {Strong}, I.~B. and {Olson}, R.~A.}]{klebesadel_1973}
   {Klebesadel}, R.~W.,  {Strong}, I.~B. \& {Olson}, R.~A.	 1973,  \apjl, 182, 85

 \bibitem[{{Knop} {et~al.}(2003) {Knop}, R.~A. and {Aldering}, G. and {Amanullah}, R. and {Astier}, P. and 
	{Blanc}, G. and {Burns}, M.~S. and {Conley}, A. and {Deustua}, S.~E. and 
	{Doi}, M. and {Ellis}, R. and {Fabbro}, S. and {Folatelli}, G. and 
	{Fruchter}, A.~S. and {Garavini}, G. and {Garmond}, S. and {Garton}, K. and 
	{Gibbons}, R. and {Goldhaber}, G. and {Goobar}, A. and {Groom}, D.~E. and 
	{Hardin}, D. and {Hook}, I. and {Howell}, D.~A. and {Kim}, A.~G. and 
	{Lee}, B.~C. and {Lidman}, C. and {Mendez}, J. and {Nobili}, S. and 
	{Nugent}, P.~E. and {Pain}, R. and {Panagia}, N. and {Pennypacker}, C.~R. and 
	{Perlmutter}, S. and {Quimby}, R. and {Raux}, J. and {Regnault}, N. and 
	{Ruiz-Lapuente}, P. and {Sainton}, G. and {Schaefer}, B. and 
	{Schahmaneche}, K. and {Smith}, E. and {Spadafora}, A.~L. and 
	{Stanishev}, V. and {Sullivan}, M. and {Walton}, N.~A. and {Wang}, L. and 
	{Wood-Vasey}, W.~M. and {Yasuda}, N.}]{knop_2003}
	{Knop}, R.~A.,  {et~al.}    2003, \apj, 598, 102 
	

\bibitem[{{Koekemoer}  {et~al.}(2011)  {Koekemoer}, A.~M. and {Faber}, S.~M. and {Ferguson}, H.~C. and 
	{Grogin}, N.~A. and {Kocevski}, D.~D. and {Koo}, D.~C. and {Lai}, K. and 
	{Lotz}, J.~M. and {Lucas}, R.~A. and {McGrath}, E.~J. and {Ogaz}, S. and 
	{Rajan}, A. and {Riess}, A.~G. and {Rodney}, S.~A. and {Strolger}, L. and 
	{Casertano}, S. and {Castellano}, M. and {Dahlen}, T. and {Dickinson}, M. and 
	{Dolch}, T. and {Fontana}, A. and {Giavalisco}, M. and {Grazian}, A. and 
	{Guo}, Y. and {Hathi}, N.~P. and {Huang}, K.-H. and {van der Wel}, A. and 
	{Yan}, H.-J. and {Acquaviva}, V. and {Alexander}, D.~M. and 
	{Almaini}, O. and {Ashby}, M.~L.~N. and {Barden}, M. and {Bell}, E.~F. and 
	{Bournaud}, F. and {Brown}, T.~M. and {Caputi}, K.~I. and {Cassata}, P. and 
	{Challis}, P.~J. and {Chary}, R.-R. and {Cheung}, E. and {Cirasuolo}, M. and 
	{Conselice}, C.~J. and {Roshan Cooray}, A. and {Croton}, D.~J. and 
	{Daddi}, E. and {Dav{\'e}}, R. and {de Mello}, D.~F. and {de Ravel}, L. and 
	{Dekel}, A. and {Donley}, J.~L. and {Dunlop}, J.~S. and {Dutton}, A.~A. and 
	{Elbaz}, D. and {Fazio}, G.~G. and {Filippenko}, A.~V. and {Finkelstein}, S.~L. and 
	{Frazer}, C. and {Gardner}, J.~P. and {Garnavich}, P.~M. and 
	{Gawiser}, E. and {Gruetzbauch}, R. and {Hartley}, W.~G. and 
	{H{\"a}ussler}, B. and {Herrington}, J. and {Hopkins}, P.~F. and 
	{Huang}, J.-S. and {Jha}, S.~W. and {Johnson}, A. and {Kartaltepe}, J.~S. and 
	{Khostovan}, A.~A. and {Kirshner}, R.~P. and {Lani}, C. and 
	{Lee}, K.-S. and {Li}, W. and {Madau}, P. and {McCarthy}, P.~J. and 
	{McIntosh}, D.~H. and {McLure}, R.~J. and {McPartland}, C. and 
	{Mobasher}, B. and {Moreira}, H. and {Mortlock}, A. and {Moustakas}, L.~A. and 
	{Mozena}, M. and {Nandra}, K. and {Newman}, J.~A. and {Nielsen}, J.~L. and 
	{Niemi}, S. and {Noeske}, K.~G. and {Papovich}, C.~J. and {Pentericci}, L. and 
	{Pope}, A. and {Primack}, J.~R. and {Ravindranath}, S. and {Reddy}, N.~A. and 
	{Renzini}, A. and {Rix}, H.-W. and {Robaina}, A.~R. and {Rosario}, D.~J. and 
	{Rosati}, P. and {Salimbeni}, S. and {Scarlata}, C. and {Siana}, B. and 
	{Simard}, L. and {Smidt}, J. and {Snyder}, D. and {Somerville}, R.~S. and 
	{Spinrad}, H. and {Straughn}, A.~N. and {Telford}, O. and {Teplitz}, H.~I. and 
	{Trump}, J.~R. and {Vargas}, C. and {Villforth}, C. and {Wagner}, C.~R. and 
	{Wandro}, P. and {Wechsler}, R.~H. and {Weiner}, B.~J. and {Wiklind}, T. and 
	{Wild}, V. and {Wilson}, G. and {Wuyts}, S. and {Yun}, M.~S.
	}]{koekemoer_2011}
         {Koekemoer}, A.~M.,      {et~al.}   2011, \apjs, 197, 36
         
 

\bibitem[{{Kulkarni}    {et~al.}(1998)  {Kulkarni}, S.~R. and {Frail}, D.~A. and {Wieringa}, M.~H. and 
	{Ekers}, R.~D. and {Sadler}, E.~M. and {Wark}, R.~M. and {Higdon}, J.~L. and 
	{Phinney}, E.~S. and {Bloom}, J.~S.}]{kulkarni_1998} 
	{Kulkarni}, S.~R.,    {et~al.}    1998, \nat, 395, 663 
	
	
\bibitem[{{Li} \&  {Hjorth}(2014) {Li}, X. and {Hjorth}, J.}] {lixue_2013sn}
                 {Li}, X.,  \& {Hjorth}, J.  2014,  ArXiv e-prints
	

	
\bibitem[{{Li}   {et~al.}(2012)  {Li}, X. and {Hjorth}, J. and {Richard}, J.}]{lenspaper}
                {Li}, X.,  {Hjorth}, J.,   \& {Richard}, J.  2012,  \jcap, 11, 15
               
\bibitem[{ {Linder} \& {Huterer} (2003)  {Linder}, E.~V. and {Huterer}, D.}]{linder_2003}
              {Linder}, E.~V., \& {Huterer}, D. 2003,  \prd, 67, 081303
              
              
              
\bibitem[{ {Metzger}    {et~al.}(1997)    {Metzger}, M.~R. and {Djorgovski}, S.~G. and {Kulkarni}, S.~R. and 
	{Steidel}, C.~C. and {Adelberger}, K.~L. and {Frail}, D.~A. and 
	{Costa}, E. and {Frontera}, F.}]{metzger_1997}
              {Metzger}, M.~R.,  {Djorgovski}, S.~G.,  {Kulkarni}, S.~R.,   
	{Steidel}, C.~C.,   {Adelberger}, K.~L.,   {Frail}, D.~A.,   
	{Costa}, E.,  \&  {Frontera}, F.  1997, \nat,  387, 878-880

 

\bibitem[{{Nasonova}  \& {Karachentsev}(2011)    {Nasonova}, O.~G. and {Karachentsev}, I.~D.}]{Nas2011} 
                {Nasonova}, O.~G.,  \& {Karachentsev}, I.~D 2011,  Astrophysics,  54, 1
                

\bibitem[{{Pan}  \&  {Loeb}(2013)  {Pan}, T. and {Loeb}, A.}]{pan_2013}
            {Pan}, T.,  \& {Loeb}, A. 2013,   \mnras, 435, 33
  
  
  
\bibitem[{{Perlmutter}  {et~al.}(1997)  {Perlmutter}, S. and {Gabi}, S. and {Goldhaber}, G. and {Goobar}, A. and 
	{Groom}, D.~E. and {Hook}, I.~M. and {Kim}, A.~G. and {Kim}, M.~Y. and 
	{Lee}, J.~C. and {Pain}, R. and {Pennypacker}, C.~R. and {Small}, I.~A. and 
	{Ellis}, R.~S. and {McMahon}, R.~G. and {Boyle}, B.~J. and {Bunclark}, P.~S. and 
	{Carter}, D. and {Irwin}, M.~J. and {Glazebrook}, K. and {Newberg}, H.~J.~M. and 
	{Filippenko}, A.~V. and {Matheson}, T. and {Dopita}, M. and 
	{Couch}, W.~J. and {Supernova Cosmology Project}}] {perlmutter_1997}
	{Perlmutter}, S., {et~al.}   1997,  \apj, 483,  565	
         

\bibitem[{{Perlmutter}  {et~al.}(1999)  {Perlmutter}, S. and {Aldering}, G. and {Goldhaber}, G. and 
	{Knop}, R.~A. and {Nugent}, P. and {Castro}, P.~G. and {Deustua}, S. and 
	{Fabbro}, S. and {Goobar}, A. and {Groom}, D.~E. and {Hook}, I.~M. and 
	{Kim}, A.~G. and {Kim}, M.~Y. and {Lee}, J.~C. and {Nunes}, N.~J. and 
	{Pain}, R. and {Pennypacker}, C.~R. and {Quimby}, R. and {Lidman}, C. and 
	{Ellis}, R.~S. and {Irwin}, M. and {McMahon}, R.~G. and {Ruiz-Lapuente}, P. and 
	{Walton}, N. and {Schaefer}, B. and {Boyle}, B.~J. and {Filippenko}, A.~V. and 
	{Matheson}, T. and {Fruchter}, A.~S. and {Panagia}, N. and {Newberg}, H.~J.~M. and 
	{Couch}, W.~J. and {The Supernova Cosmology Project}}]{perlmutter}
	{Perlmutter}, S.,    {et~al.}     1999,   \apj, 517,  565

	




\bibitem[{{Phillips}(1993) {Phillips}, M.~M.}]{phillips_1993}
                  {Phillips}, M.~M., 1993, \apjl, 413, 105
                  
     
\bibitem[{{Phillips}  {et~al.}(1999)   {Phillips}, M.~M. and {Lira}, P. and {Suntzeff}, N.~B. and {Schommer}, R.~A. and 
	{Hamuy}, M. and {Maza}, J.}]{phillips_1999}
	{Phillips}, M.~M.,   {Lira}, P.,   {Suntzeff}, N.~B.,  {Schommer}, R.~A.,   
	{Hamuy}, M., \&  {Maza}, J.   1999, \aj, 118, 1766
	
\bibitem[{{Planck Collaboration}   {et~al.}(2013)               {Planck Collaboration} and {Ade}, P.~A.~R. and {Aghanim}, N. and 
	{Armitage-Caplan}, C. and {Arnaud}, M. and {Ashdown}, M. and 
	{Atrio-Barandela}, F. and {Aumont}, J. and {Baccigalupi}, C. and 
	{Banday}, A.~J. and et al.  }]{planck_cosmopara}
	{Planck Collaboration}   {et~al.}   2013 , ArXiv e-prints
	

\bibitem[{{Riess}    {et~al.}(1998)        {Riess}, A.~G. and {Filippenko}, A.~V. and {Challis}, P. and 
	{Clocchiatti}, A. and {Diercks}, A. and {Garnavich}, P.~M. and 
	{Gilliland}, R.~L. and {Hogan}, C.~J. and {Jha}, S. and {Kirshner}, R.~P. and 
	{Leibundgut}, B. and {Phillips}, M.~M. and {Reiss}, D. and {Schmidt}, B.~P. and 
	{Schommer}, R.~A. and {Smith}, R.~C. and {Spyromilio}, J. and 
	{Stubbs}, C. and {Suntzeff}, N.~B. and {Tonry}, J.}]{riess}
	{Riess}, A.~G.,   {et~al.}    1998,  \aj,  116,  1009
	


	
\bibitem[{{Rodney}    {et~al.}(2014)       {Rodney}, S.~A. and {Riess}, A.~G. and {Strolger}, L.-G. and 
	{Dahlen}, T. and {Graur}, O. and {Casertano}, S. and {Dickinson}, M.~E. and 
	{Ferguson}, H.~C. and {Garnavich}, P. and {Hayden}, B. and {Jha}, S.~W. and 
	{Jones}, D.~O. and {Kirshner}, R.~P. and {Koekemoer}, A.~M. and 
	{McCully}, C. and {Mobasher}, B. and {Patel}, B. and {Weiner}, B.~J. and 
	{Cenko}, S.~B. and {Clubb}, K.~I. and {Cooper}, M. and {Filippenko}, A.~V. and 
	{Frederiksen}, T.~F. and {Hjorth}, J. and {Leibundgut}, B. and 
	{Matheson}, T. and {Nayyeri}, H. and {Penner}, K. and {Trump}, J. and 
	{Silverman}, J.~M. and {U}, V. and {Azalee Bostroem}, K. and 
	{Challis}, P. and {Rajan}, A. and {Wolff}, S. and {Faber}, S.~M. and 
	{Grogin}, N.~A. and {Kocevski}, D.}] {rodney_snr}
         {Rodney}, S.~A.,    {et~al.}   2014, \aj,  148,  13
         



\bibitem[{ {Schlafly}    {et~al.}(2011)    {Schlafly}, E.~F. and {Finkbeiner}, D.~P.}]{schlafly_dustmap}
               {Schlafly}, E.~F., \& {Finkbeiner}, D.~P. 2011, \apj, 737, 103
               
      
\bibitem[{ {Schlegel}     {et~al.}(1998)      {Schlegel}, D.~J. and {Finkbeiner}, D.~P. and {Davis}, M.}] {schlegel_dustmap}
                    {Schlegel}, D.~J.,  {Finkbeiner}, D.~P., \&  {Davis}, M.    1998,  \apj,  500,   525
                    

\bibitem[{{Spergel}     {et~al.}(2003)         {Spergel}, D.~N. and {Verde}, L. and {Peiris}, H.~V. and {Komatsu}, E. and 
	{Nolta}, M.~R. and {Bennett}, C.~L. and {Halpern}, M. and {Hinshaw}, G. and 
	{Jarosik}, N. and {Kogut}, A. and {Limon}, M. and {Meyer}, S.~S. and 
	{Page}, L. and {Tucker}, G.~S. and {Weiland}, J.~L. and {Wollack}, E. and 
	{Wright}, E.~L.}]{spergel_2003}
	{Spergel}, D.~N.,   {et~al.}   2003,   \apjs,  148, 175
	
	
\bibitem[{{Stanek}    {et~al.}(2003)                     {Stanek}, K.~Z. and {Matheson}, T. and {Garnavich}, P.~M. and 
	{Martini}, P. and {Berlind}, P. and {Caldwell}, N. and {Challis}, P. and 
	{Brown}, W.~R. and {Schild}, R. and {Krisciunas}, K. and {Calkins}, M.~L. and 
	{Lee}, J.~C. and {Hathi}, N. and {Jansen}, R.~A. and {Windhorst}, R. and 
	{Echevarria}, L. and {Eisenstein}, D.~J. and {Pindor}, B. and 
	{Olszewski}, E.~W. and {Harding}, P. and {Holland}, S.~T. and 
	{Bersier}, D.}] {stanek_2003dh}
	{Stanek}, K.~Z.,  {et~al.}   2003,  \apjl,  591, 17
	


\bibitem[{{Woosley}  \&   {Bloom}(2006)   {Woosley}, S.~E. and {Bloom}, J.~S.}] {woosley_2006}
                 {Woosley}, S.~E., \& {Bloom}, J.~S.  2006,  \araa,  44, 507




\bibitem[{{Woosley}    {et~al.}(1999)        {Woosley}, S.~E. and {Eastman}, R.~G. and {Schmidt}, B.~P.}]{woosley_1998}
                 {Woosley}, S.~E., \& {Eastman}, R.~G. and {Schmidt}, B.~P.  1999, \apj,  516, 788
                 

\end{thebibliography}
  \end{document}